\documentclass[journal]{IEEEtran}

\usepackage{ifpdf}
\usepackage{cite}

\ifCLASSINFOpdf
  \usepackage[pdftex]{graphicx}
  \graphicspath{{../Figures/}}
  \DeclareGraphicsExtensions{.pdf,.jpeg,.png}
\else
 \usepackage[dvips]{graphicx}
  \graphicspath{{../Figures/}}
  \DeclareGraphicsExtensions{.eps,.ps}
\fi

\usepackage[cmex10,centertags]{amsmath}
\usepackage{amsbsy}

\usepackage{array}
\usepackage{mdwmath}
\usepackage{mdwtab}
\usepackage{amsfonts}
\usepackage{amssymb}
\usepackage{url}

\begin{document}

\title{Compact source detection \\ 
 in multi-channel microwave surveys: \\
 from SZ clusters to polarized sources}

\author{Diego~Herranz$^{1,3}$,~Francisco~Arg\"ueso$^2$~and~Pedro~Carvalho$^3$%
\thanks{$^1$ Instituto de F\'\i sica de Cantabria, CSIC-UC, Av. los Castros s/n, 39005 Santander, Spain}%
\thanks{$^2$ Departamento de Matem\'aticas, Universidad de Oviedo, Calvo Sotelo s/n, 33007 Oviedo, Spain}%
\thanks{$^3$ Astrophysics Group, Cavendish Laboratory, JJ Thomson Avenue, Cambridge CB3 0HE, UK}}

\markboth{}%
{}

\maketitle

\begin{abstract}
In this paper we describe the state-of-the art status of
multi-frequency detection techniques for compact sources in
microwave astronomy. From the simplest cases where the spectral
behaviour is well-known (i.e. thermal SZ clusters) to the more
complex cases where there is little a priori information (i.e.
polarized radio sources) we will review the main advances and the
most recent results in the detection problem.
\end{abstract}

\begin{IEEEkeywords}

\end{IEEEkeywords}

\section{Introduction}

Extragalactic foregrounds play a crucial role in microwave astronomy, not only by their effect as contaminants of the Cosmic Microwave Background (CMB) but also by their own right as cosmological probes. Galaxies and galaxy clusters, if not properly identified and taken into account, can seriously affect the measurement of the CMB anisotropies angular power spectrum in temperature~\cite{TO98,zotti05,tucci11} and polarization~\cite{tucci04,tucci05}, CMB non-Gaussianity tests~\cite{argueso03,komatsu03,smith04,barreiro06,argueso06,babich08,lacasa11} and even the performance of component separation methods used for the study of Galactic foregrounds\footnote{On the other hand, the opposite is also true: foregrounds can affect the performance of compact source detection algorithms. In general, compact source detection algorithms find it easier to deal with diffuse foregrounds than diffuse component separation techniques to deal with compact sources, so the typical CMB analysis pipeline includes the detection of compact sources as a previous step to diffuse component separation.}~\cite{vielva01,leach08}. On the other hand, galaxy and galaxy cluster surveys in the sub-mm regime of the electromagnetic spectrum are powerful tools for cosmology~\cite{carlstrom02,zotti10,ATLAS}.
This is the motivation of the considerable number of works on extragalactic foreground detection that have appeared in the literature on recent years.

As opposed to Galactic foregrounds, that are typically extended as diffuse clouds over large areas of the sky, individual extragalactic objects appear as compact blobs that subtend very small angular scales. For this reason, both galaxies and galaxy clusters are often referred to as \emph{compact sources} and their detection/separation is typically treated as a problem apart from the one posed by the separation of Galactic diffuse components.   

It is precisely the compactness of extragalactic sources that makes it possible to detect them against the fluctuations of the diffuse components (CMB included) in single-frequency (channel) settings. Most of the detection methods that have been proposed in the literature make use of this \emph{scale diversity}. The well known SExtractor package~\cite{sex}, for example, is particularly good at estimating and then subtracting the background at coarse scales and then detecting compact sources by looking for small sets of connected pixels above a given threshold. For the same reason, techniques based on scale-selecting devices such as band-pass filters~\cite{tegmark98,sanz01,herr02a,herr02b,caniego05,caniego06} and wavelets~\cite{vielva03,MHWF} have proven to be very useful. More sophisticated detection Bayesian algorithms~\cite{hobson03,feroz08,carvalho09,HobsonBayes,argueso11} make also use at some point of the scale diversity of compact sources versus diffuse foregrounds. A recent review on methods for the detection of compact sources in microwave images can be found in~\cite{herr10}. 

The majority of current CMB experiments, such as the Wilkinson Microwave Anisotropy Probe (WMAP)~\cite{map} and \emph{Planck}~\cite{Planck}, can observe the sky in several frequency bands simultaneously. Some of them are also capable of measuring not only the intensity of the microwave radiation, but also its polarization. This means that extragalactic compact sources can be observed in several different images or \emph{channels}. Multi-channel\footnote{In this work we wil use the term \emph{multi-channel detection} instead of the more common \emph{multi-frequency} term because it can accommodate both the classic problem of detecting a source using different frequency maps and the detection of a source in polarization using images of the Stokes' parameters $I$, $Q$, $U$. As we will see, both problems are formally equivalent and therefore we prefer the more general terminology.}  information can be used to improve the chances of detecting compact sources. In this paper we will review the state of the art multi-channel detection techniques for compact sources in microwave astronomy.

Multi-channel (multi-frequency) astronomy is almost as old as the telescope itself: we have been doing it since the first colour filters became available. During the XX$^{\mathrm{th}}$ and XXI$^{\mathrm{st}}$ centuries astronomy has expanded its range of operation into the radio, microwave, infrared, ultraviolet, X-ray and gamma regimes of the electromagnetic spectrum: a tremendous amount of information we are only beginning to piece together. The most basic multi-channel consideration we can do is: given that we have observed some object in the optical, let us see how does it look like in, for example, the infra-red. Catalogue cross-correlation and band-merging, follow-ups and the construction of spectral energy distribution (SED) curves are fundamental parts of this multi-channel quest for knowledge. But we are \emph{not} going to review this. 

A follow-up --looking at a certain position of the sky where a particular source has been already detected in a different region of the electromagnetic spectrum-- can only work if we have a clear detection of the source in the first place. Sometimes, however, the sources are just too faint to be detected with enough significance in any of the available channels. But maybe if we could join together all the channels we would be able to detect the source. Even in the cases where the source is clearly detected in one channel, it may be too weak to be observed in other channels. Once again, one can ask if there is any possibility to use other information apart from just the source position from the `good' channels in order to enhance the source in the 'bad' ones. Or we can more generally ask if there is a better way to combine all the channels to get better SEDs than just going channel by channel separately. This is the aim of our review. 

The key for multi-channel source separation is \emph{spectral diversity}: we hope the signal of interest to scale from one image to other in a different way than the other components (background). Assuming a linear mixture model in which $N$ different, unknown sources are added with a set of channel-dependent weights to produce $M$ observed channels ($M \geq N$), and making certain assumptions about the statistical properties of the sources --for example statistical independence, or non-Gaussianity, etc-- the Blind Source Separation (BSS) problem is solvable by means of a number of statistical signal processing techniques such as the Maximum Entropy Method~\cite{MEM}, Independent Component Analysis~\cite{ICA}, Correlated Component Analysis~\cite{CCA1,CCA2} or wavelet-based Internal Template Fitting~\cite{WIFIT}, just to mention a few of them. For a more detailed review of BSS methods applied to microwave experiments, see~\cite{leach08}.

Unfortunately for our purposes the above mentioned BSS techniques are not well suited for  the detection of extragalactic compact sources, except  for the particular case of galaxy clusters observed through the thermal Sunyaev-Zel'dovich (tSZ) effect. Individual galaxies   
leave their imprint on the microwave sky through an enormous variety of astrophysical mechanisms --from radio active lobes to dust thermal emission-- so that, strictly speaking, each individual galaxy has its own unique spectral behaviour. In the linear mixing model this translates into $N \gg M$ and the BSS problem is sorely under determined. New methods, specifically tailored for compact sources, become necessary.
Even in the case of tSZ clusters, where all the sources share the same spectral behaviour, it may be advisable to apply other techniques different from the above mentioned BSS methods. The tSZ is sub-dominant at all frequencies, making it very difficult for BSS techniques to detect but the brightest clusters in the sky. The multi-channel detection methods we are going to describe in this review make use, up to different extent, of both \emph{scale} and {spectral} diversities in order to optimize the detectability of compact sources.

In this paper we will review the most widely used methods for detecting extragalactic compact sources using multi-channel data. We will first formalize the problem in section~\ref{sec:model}. Then we will review the different approaches currently used in microwave astronomy. We choose the order in which we introduce the methods on the basis of the amount of information that is used to achieve detection: we will start in section~\ref{sec:stack} with traditional stacking and band-merging techniques that make a minimal use of multi-channel information and then proceed in section~\ref{sec:mtxf} to a linear filtering scheme that takes into account the correlation of the background among different channels. In section~\ref{sec:mmf} we will discuss how to filter the data when both the correlation of the background among channels and the spectral behaviour of the sources are known. We will see how the second condition can be relaxed. Then we will discuss the more general theory of Bayesian detection in section~\ref{sec:bayes}. Finally, we will devote section~\ref{sec:pol} to the particular case of polarization data.

\section{Compact sources in multi-channel observations} \label{sec:model}

Let us consider a set of $N$ two-dimensional images (channels) in which there is an unknown number of compact sources embedded in a mixture of instrumental noise and other astrophysical components. Without loss of generality, let us consider the case of a single source with a certain typical angular scale $R$ and located at the origin of the coordinates. Our data model is
\begin{equation} \label{eq:model1}
D_j\left(\vec{x}\right) = s_j\left(\vec{x};R\right) + n_j\left(\vec{x}\right),
\end{equation}
\noindent
where the subscript $j = 1, \ldots , N$ denotes the index of the channel: it may refer to a given frequency (i.e. 30 GHz, 44 GHz, etc), to a polarization channel (i.e. the Stokes' parameters $I,Q,U$) or any other image indexing we can consider. The scale $R$ of the sources can vary from one object to other (as it happens for galaxy clusters) or be the same for all the objects belonging to a given class (i.e. radio sources observed by low angular resolution experiments such as WMAP). It is common to factorize the source term as the product of a channel-dependent \emph{amplitude} or \emph{intensity} and a spatial profile
\begin{equation} \label{eq:model2}
s_j\left(\vec{x}\right) = A_j \times \tau_j\left(\vec{x};R\right).
\end{equation}
\noindent
The spatial profile, in turn, includes the possible effects of any channel-dependent point spread function (beam):
\begin{equation} \label{eq:model3}
\tau_j\left(\vec{x};R\right) = b_j\left(\vec{x}\right) \otimes \tau^0_j\left(\vec{x};R\right).
\end{equation}
\noindent
The above formula can be expressed more easily in Fourier space:
\begin{equation} \label{eq:model4}
\tau_j\left(\vec{k};R\right) = b_j\left(\vec{k}\right) \times \tau^0_j\left(\vec{k};R\right),
\end{equation}
\noindent
where for simplicity we have kept the same symbol for the functions in real and in Fourier space (the argument $\vec{x}$ or $\vec{k}$ indicates clearly enough in which space we are).
We choose the normalization of the profile so that
\begin{equation} \label{eq:model5}
\tau^0_j\left(\vec{0};R\right)=1.
\end{equation}
\noindent
Finally, for simplicity we will consider symmetric beams and profiles, $\tau_j\left(\vec{x};R\right) = \tau_j \left(|\vec{x}|;R\right) = \tau_j (x;R)$. This assumption is not strictly necessary and can be relaxed on demand, but it greatly simplifies calculations and is quite reasonable in most applications.

The term $n_j\left(\vec{x}\right)$ in equation (\ref{eq:model1}) is the generalized noise in the $j^{\mathrm{th}}$ channel, containing not only instrumental noise, but also CMB and all the other astrophysical components apart from the compact sources. Let us suppose that the noise has zero mean and that it can be characterized by its cross-power spectrum:
\begin{equation} \label{eq:PS}
\langle n_j\left(\vec{k}\right) n^*_{l}\left(\vec{k^{\prime}}\right) \rangle =
P_{jl}\left(k\right) \delta^2 \left(\vec{k}-\vec{k^{\prime}}\right).
\end{equation}
In other words, we work under the assumption that the properties of the noise can be sufficiently described by its second-order statistics. Please note that the homogeneity of the background is a necessary condition for the power spectrum to be a full second order description of the background; this condition is not globally met in real astronomical images (where the Galactic foregrounds, for example, are strongly non homogeneous), but we can always take patches small enough to satisfy local homogeneity (at least on a first approximation).

\subsection{Notation}

In this paper we will use the notation $\vec{x}$ to indicate coordinate vectors (both in real and Fourier spaces) and the notation $\mathbf{x}$ to indicate vectors whose components are labelled with the indexes of the different channels. According to this notation, we can write equations (\ref{eq:model1}--\ref{eq:PS}) as
\begin{eqnarray}
\mathbf{D}\left(\vec{x}\right) & = & \mathbf{s}\left({x};R\right) + \mathbf{n}\left(\vec{x}\right) \nonumber \\
\mathbf{s}\left({x}\right) & = & \mathbf{A} \boldsymbol{\tau}\left({x};R\right) \nonumber \\
\boldsymbol{\tau}\left({k};R\right) & = & \mathbf{b}\left({k}\right) \mathbf{\tau^0}\left({k};R\right) \nonumber \\
\langle \mathbf{n}^t \left( \vec{k} \right) \mathbf{n} \left( \vec{k^{\prime}} \right) \rangle & = & \mathbf{P} (k) \delta^2 \left( \vec{k} - \vec{k^{\prime}} \right).
\end{eqnarray}
It will be useful in some cases to expand the vector of intensities as
\begin{equation} \label{eq:amplitud}
\mathbf{A} = I_* \mathbf{f},
\end{equation}
\noindent
where $I_*$ can be interpreted as a channel-independent intensity and $\mathbf{f}$ is a vector containing the spectral behaviour of the sources across the different channels. For example, when we consider the tSZ $I_*$ can be associated with the cluster Compton-$y$ parameter and $\mathbf{f}$ takes the well-known form, in thermodynamic temperature units,
\begin{equation} \label{eq:tSZ}
f_{\hat{\nu}} \propto  \hat{\nu} \frac{e^{\hat{\nu}}+1}{e^{\hat{\nu}}-1} - 4 
\end{equation}
\noindent
where $\hat{\nu} = h \nu / kT$, $\nu$ is the frequency of observation, $k$ is the Boltzmann constant and $T$ is the temperature of the CMB. Other example is provided by radio sources whose spectral behaviour can be approximated by a power law
\begin{equation} \label{eq:powlaw}
f_{\nu} \propto \left( \frac{\nu}{\nu_0} \right)^{-\alpha},
\end{equation}
\noindent
where $\nu_0$ is some frequency of reference and $\alpha$ is known as the \emph{spectral index}. In this case, $I_*$ can be interpreted as the source flux density at the reference frequency $\nu_0$. Please note that analytic spectral laws such as (\ref{eq:tSZ}) or (\ref{eq:powlaw}) are not always available, or even convenient.

A list of symbols used in this paper can be found in table~\ref{tb:list}, that appears in the Appendix.

\section{Basic multi-channel detection} \label{sec:stack}

Probably, the simplest imaginable approach to the multi-channel problem consists in trying somehow to transform it to a much simpler single-channel problem.  The theory and applications of single-channel source detection are well studied in the literature; we will assume in this work that the reader is already familiar with the topic. A short, recent review can be found in \cite{herr10} and references therein. 

Let us start with the simplest --and most frequent-- case of multi-channel observation we can conceive. Imagine we have just two independent observations $D_1$ and $D_2$ of the same source, taken at the same frequency and with the same instrumental characteristics (beam size, noise level, etc). A perfect example is any pair of identical radiometers in an experiment such as \emph{Planck}. Let us also assume that the two observations are simultaneous, or at least that the source under study has not experienced variability in the time passed between observations. Then, it is well known that if the noises of the two observations are Gaussian with rms $\sigma_1=\sigma_2=\sigma$ and independently distributed, the linear combination {$\tilde{D}=(D_1+D_2)/2$} has a lower rms noise $\tilde{\sigma}=\sigma/\sqrt{2}$. If the different channels have different noise levels but the noises are still uncorrelated, another well-known result is that the noise of the linear combination of the channels can be minimized by inverse noise variance weighting of the individual channels. Once the multi-channel data has been combined (or \emph{projected}) into a single channel (or \emph{plane}) it is straightforward to apply threshold detection, or filtering plus thresholding, or any other of the techniques described in~\cite{herr10}. With the appropriate modifications, the weighting internal linear combination scheme can be extended to correlated noise~\cite{naselsky02}. The problem with this approach is that if the source has not the same intensity in all the channels (or if its spectral behaviour is not perfectly known) it is impossible to relate the intensity observed in the combined image with the true intensity of the source.

Other merging schemes can be advisable in some other circumstances. Imagine that some of the components of the noise in (\ref{eq:model1}) are constant across all the channels: the best example is the CMB itself in images expressed in thermodynamic temperature units. Then an obvious way to reduce the noise is to subtract pairs of channels. As an example, \cite{chen08} and \cite{wright09} used combinations of the WMAP W and V bands in order to produce a CMB-free map to better detect the elusive radio galaxies in the WMAP 5-year data. The problem with this approach is that if a source has by chance a spectrum flat enough, it may be also cancelled by the subtraction.

\section{Matrix Multi-Filters} \label{sec:mtxf}

\begin{figure*}[!t]
\centering
\includegraphics[width=\textwidth]{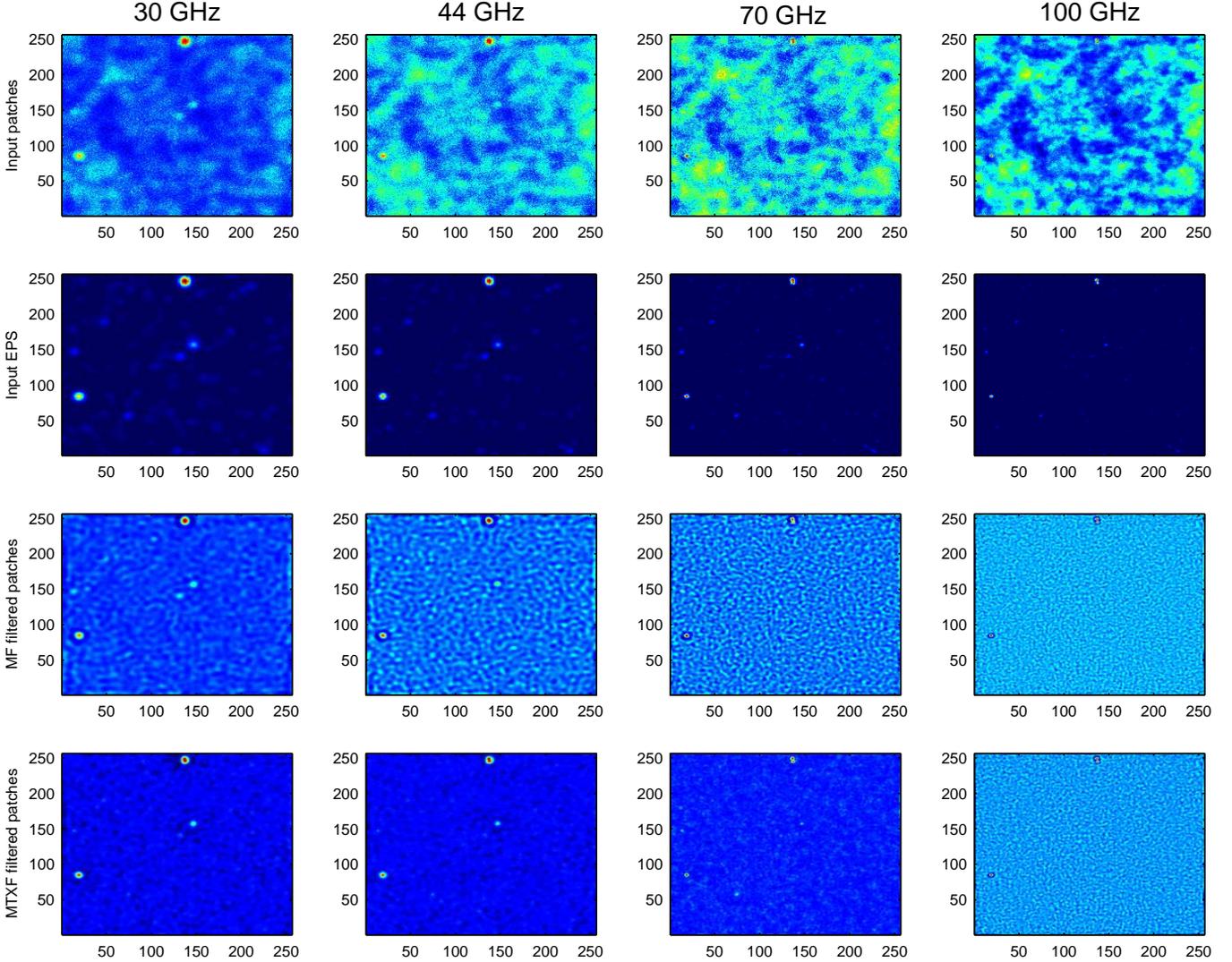}
\caption{Comparison between single-channel matched filtering and multi-channel MTXF filtering for a four-channel simulation of a patch of the sky as observed by the \emph{Planck} 30, 44, 70 and 100 GHz detectors. The top row of images show the simulated patches. The second row shows just the point sources present in the simulations. The third row shows the patches after being filtered with single-channel matched filters. The fourth row shows the patches filtered with the corresponding MTXF. The $x$ and $y$-axes are in pixel units (pixel size = 3.435 arcmin) }
\label{fig:MTXF}
\end{figure*}

A recurrent problem of channel linear combination is that accurate photometry --that is, the determination of $I_*$-- is not possible unless the exact values of all the components of the spectral behaviour $\mathbf{f}$ are known. But the spectral behaviour is not known in most cases. General laws such as (\ref{eq:powlaw}) are not reliable when the range of frequencies involved is wide enough, and even if they are there is the additional problem that different sources in a given patch of the sky will have different spectral indexes. The classic solution in these cases, if one wants to persist using linear combination nevertheless, is to go back to the individual channels and extract photometric measurements on them.

 In order to overcome these problems \cite{herr08,herr09} proposed a method that combines the signal-to-noise boosting capacity of linear filters with a particular merging scheme that makes totally irrelevant the spectral behaviour $\mathbf{f}$. The method is called \emph{matrix multi-filters} (MTXF) and its foundations can be summarized as follows:

Let $\Psi_{ij} \left( \vec{x} \right)$ be a set (matrix) of $N \times N$  linear filters, and let us define the set of quantities
\begin{equation} \label{eq:mtxf_field}
w_i \left( \vec{x} \right) = \sum_ j \int d \vec{y} \ \Psi_{ij} \left( \vec{x} - \vec{y} \right) D_j \left( \vec{y} \right).
\end{equation}

The quantity above is the sum of a set of linear filterings of the individual channels. 
We intend to use the combined filtered images $w_i\left( \vec{x} \right)$ as estimators of the source amplitudes $A_i$. For that purpose, we require that the filters $\Psi_{ij}$ satisfy the conditions:
\begin{itemize}
\item The combined filtered image at the position of the source $w_i\left( \vec{0} \right)$ is an \emph{unbiased} estimator of $A_i$.
\item The combined filtered image at the position of the source $w_i\left( \vec{0} \right)$ is an \emph{efficient} estimator of $A_i$, that is, the variance of the estimator is minimum.
\end{itemize} 
Looking at the structure of (\ref{eq:mtxf_field}) it is easy to verify that a sufficient condition in order to guarantee the first condition above is
\begin{equation} \label{eq:ortho}
\int d \vec{y}  \ \Psi_{ij} \left( \vec{x} - \vec{y} \right) \tau_j \left( \vec{y} \right) = \delta_{ij} \left( \vec{x} \right),
\end{equation}
\noindent
that is, the filter functions are orthonormal to the source profile functions. This guarantees statistical unbiasedness independently of whatever values $\mathbf{f}$ may take. The solution to the problem given the two conditions above and the constraints (\ref{eq:ortho}) is, in Fourier space
\begin{equation} \label{eq:mtxf}
\mathbf{\Psi^*} = \boldsymbol{\Lambda}\mathbf{P}^{-1},
\end{equation}
\noindent
where $*$ denotes complex conjugation and 
\begin{eqnarray}\label{eq:mtxf2}
\Lambda_{lm}     & = & \lambda_{lm} \tau_m \nonumber \\
\boldsymbol{\lambda} & = & \mathbf{H}^{-1} \nonumber \\
H_{lm}           & = & \int d \vec{k} \ \tau_l \left( \vec{k} \right) P_{lm}^{-1} \left({k} \right) \tau_m^* \left( \vec{k} \right) . 
\end{eqnarray}
In the case where $\mathbf{P}$ is diagonal (no noise correlation among channels), it can be shown that the matrix multi-filters default to a simple matched filter applied individually to each channel. 

\begin{figure}[!t]
\centering
\includegraphics[width=\columnwidth]{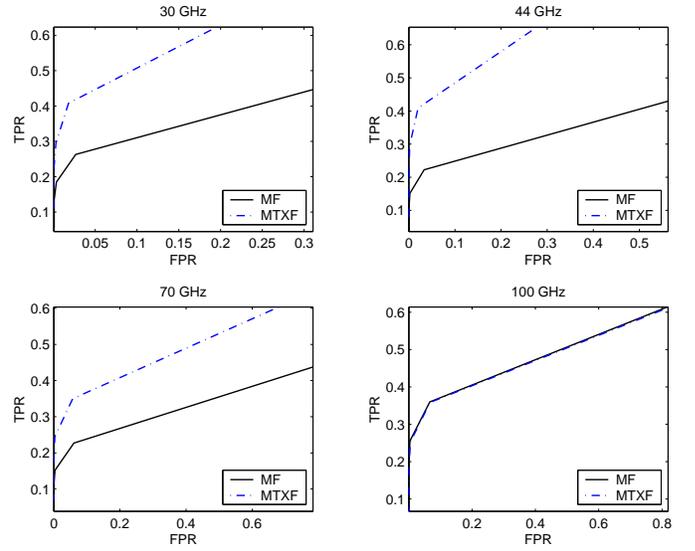}
\caption{ROC curves for the filtering schemes and the four channels considered.
Solid line: MF. Dot-dashed line: matrix filters. Axis labels: TPR stands for True Positives Ratio (associated to the completeness of a catalogue), FPR stands for False Positives Ratio (associated to the purity of the catalogue, a low FPR indicates a high purity).  }
\label{fig:ROC}
\end{figure}

The MTXF do not project $N$ channels into a single effective plane where to detect the sources. They project instead $N$ channels into $N$ new planes where the sources can be detected and their $N$ amplitudes $A_i$ $(i=1,\ldots,N)$ can be estimated separately. The difference with the single-channel approach is that each one of the $N$ new planes $w_i$ is constructed by combination of the original $N$ channels in such a way the noise is cancelled more effectively: the MTXF use the multi-channel cross-correlation of noise but do not use at all any information about the spectral behaviour of the sources. It can be described as a `semi multi-channel approach', in the sense that it uses only half of the available information. When the noise cross-correlation among channels is zero, the method defaults to standard single-channel matched filtering. But when the cross-correlation is not null, the MTXF give better signal-to-noise boosts than the standard single-channel matched filter. Figure~\ref{fig:MTXF} shows the comparison between the single-channel matched filter and the MTXF for a simulation of the \emph{Planck} 30, 44, 70 and 100 GHz channels.  Note that for the 44 and 70 GHz channels the output matrix filtered maps look far cleaner than their matched filtered equivalents. For the 30 GHz channel, the distinction is not so clear, but some improvement can be appreciated nevertheless. Finally, for the 100 GHz channel both filtered images look practically identical. The gain factors obtained for these images with the MTXF are [2.9, 3.8, 3.5, 2.8] for the [30, 44, 70, 100] GHz channels. The signal-to-noise gain ratio between the MTXF and the matched filter is GMTXF/GMF = [1.38, 1.52, 1.49, 1.00] for the [30, 44, 70, 100] GHz channels. Note that in one of the channels (100 GHz) the signal-to-noise gain is equal to one; although this is not always the case, the MTXF tend to default to the standard matched filter for some of the channels when the multi-channel mixing does not add 
information in a constructive way: this is often (but not always) the case for the \emph{Planck} 100 GHz when combined with the lower frequency channels, that have worse angular resolution and higher noise levels. This cannot be taken as a general rule, because the conditions change from one region of the sky to another. Only in \emph{the worst} case the 100 GHz MTXF filtered image is as good as the single-channel MF filtered image.  Regarding the number of true and spurious detections produced by both methods Figure~\ref{fig:ROC} shows the receiver operating characteristic (ROC) curves for the MTXF and the MF in the four considered channels;
a clear improvement can be appreciated at 30, 44 and 70 GHz, whereas both methods work similarly for thr 100 GHz case.
This serves as an indication of how MTXF can produce results better or at least equal to single-channel matched filters without making any use of the sources spectral diversity.

\section{Matched Multi-Filter} \label{sec:mmf}

Multi-channel detection techniques reach the summit of their power when they fully exploit the spectral diversity of compact sources with respect to the background. 
he fully multi-channel compact source detection techniques that are being applied in the context of CMB astronomy lie in two main groups: linear filtering methods based on the so-called matched multi-filters plus some thresholding detection criteria and Bayesian methods, to be discussed in the next section.

\subsection{Standard matched multi-filters}

Let us assume that the spectral behaviour $f$ of the sources is known \emph{a priori}. An example is the well-known thermal Sunyaev-Zel'dovich effect which, ignoring relativistic effects, has a spectral behaviour given by (\ref{eq:tSZ}). This opens interesting possibilities. For example, if $\mathbf{f}$ is known, it is straightforward to stack the different channels with an optimal weighting designed to minimize the effects of the background while keeping intact the intensity of the sources \cite{herr02c}:
\begin{equation} \label{eq:ILCSZ}
\tilde{D} \left(\vec{x}\right) = \sum_i w_i D_i \left(\vec{x}\right) = \mathbf{w}^t \mathbf{D}\left(\vec{x}\right),
\end{equation}
\noindent
such that if there is a source of intensity $I_*$ located at the origin, then
\begin{equation}
\langle \tilde{D}\left(\vec{0}\right) \rangle = I_*.
\end{equation}
The solution to this kind of internal linear combination of channels is given by the generalized eigenvalue problem\cite{herr02c}
\begin{equation} \label{eq:ILCSZ2}
\left( \mathbf{G}-\lambda \mathbf{M} \right) \mathbf{w} = 0,
\end{equation}
\noindent
where the elements of the matrices $\mathbf{G}$ and $\mathbf{M}$ are given by
\begin{eqnarray}
G_{ij} & = & f_i \tau_i (0) \tau_j (0) f_j , \\
M_{ij} & = & \langle n_i \left(\vec{x}\right) n_j \left(\vec{x}\right)  \rangle .  
\end{eqnarray}
\noindent
It is evident that $\mathbf{M}$ is a measure of the cross-correlation of the noise among the different channels. The combination (\ref{eq:ILCSZ}) using the weights obtained through (\ref{eq:ILCSZ2}) gives the optimal internal linear combination (ILC) map for the detection of sources with the spectral behaviour $\mathbf{f}$. The ILC map can be further processed using a single-channel matched filter as suggested by~\cite{herr02c}. In an independent work, \cite{naselsky02} combined simulated multiwavelength maps in order to increase the average signal-to-noise ratio of galaxies assuming that their spectral behaviour could be modelled by expressions such as (\ref{eq:powlaw}) with a known spectral index.

In the previous ILC method, however, the separation between linear combination and filtering seems somewhat artificial. The process of filtering and projecting into a single effective plane can be achieved in a single step by means of the so-called \emph{matched multi-filters}~\cite{herr02c,herr05,melin06,schaf06}. Let us define the effective filtered plane as a sum of optimally filtered images
\begin{equation} \label{eq:MMF1}
\hat{D} \left( \vec{x}; S \right) = \sum_i \int d\vec{y} \ D_i\left(\vec{y}\right) \psi_i \left( \vec{x} - \vec{y}; S \right),
\end{equation}
\noindent
where the $\psi_i$ are $N$ linear filters depending on a certain scale parameter $S$.  The meaning of this scale parameter will become evident shortly. If we impose to the effective filtered plane $\hat{D}$ the usual unbiasedness and efficiency conditions, that is
\begin{itemize}
\item At the position of the sources, $\hat{D}$ is an unbiased estimator of $I_*$.
\item The variance of $\hat{D}$ is minimum.
\end{itemize}
The solution, as proven in~\cite{herr02c}, is given in Fourier space by
\begin{eqnarray} \label{MMF}
\mathbf{\psi} \left(\vec{k};S\right) & = & \alpha(S) \ \mathbf{P}^{-1}\left(\vec{k}\right) \mathbf{F}\left(\vec{k};S\right), \\ 	
\alpha^{-1}(S) & = & \int d \vec{k} \ \mathbf{F}^t\left(\vec{k};S\right) \mathbf{P}^{-1}\left(\vec{k}\right) \mathbf{F}\left(\vec{k};S\right),
\end{eqnarray} 
\noindent 
where the vector $\mathbf{F}$ has components $F_i \left(\vec{k};S\right)= f_i \tau_i \left(\vec{k};S\right)$. We show explicitly the dependence on $S$ for a reason that will be clear immediately. The filters defined by (\ref{MMF}) take the name of \emph{matched multi-filters} (MMF). 

\begin{figure}[!t]
\centering
\includegraphics[width=\columnwidth]{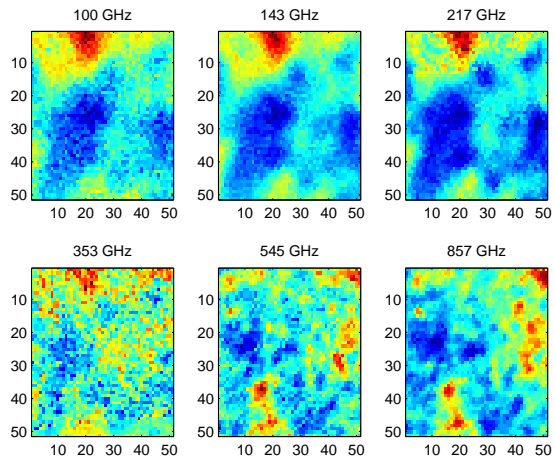}
\caption{Simulated patches of the sky at 100, 143, 217, 353, 545 and 857 GHz. A bright galaxy cluster is in the centre of the patches, but it is almost invisible among the CMB and Galaxy fluctuations.}
\label{fig:patches}
\end{figure}

\begin{figure}[!t]
\centering
\includegraphics[width=\columnwidth]{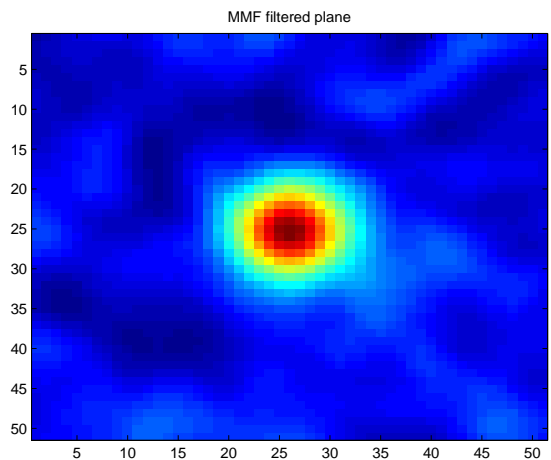}
\caption{The same area of the simulated sky as in Figure~\ref{fig:patches}, after filtering with MMF. Now the galaxy cluster is clearly visible.}
\label{fig:filtered}
\end{figure}

Just to show the power of the MMF, in Figures~\ref{fig:patches} and~\ref{fig:filtered} we show the unfiltered channels of a typical simulated SZ observations and the corresponding filtered plane. Figure~\ref{fig:patches} shows a small part of a realistic sky simulation obtained with the \emph{Planck} Sky Model (PSM)  package \cite{PSMpap} for the \emph{Planck} HFI frequencies of 100 143, 217, 353, 545 and 857 GHz. There is a bright galaxy cluster in the center of the images, but it is very hard to spot it visually. On the other hand, the presence of the cluster is evident in the MMF-filtered image shown in Figure~\ref{fig:filtered}.

\subsection{Objects with different size} \label{sec:sizes}

Until now, we have not made reference to the scale $R$ of the sources that appeared in equations (\ref{eq:model1})-(\ref{eq:model5}). In sections~\ref{sec:stack} and~\ref{sec:mtxf} we have implicitly assumed that all the sources shared the same basic profile $\tau_0$. This is pretty well the case of galaxies in low resolution experiments such as WMAP and \emph{Planck}: the angular scale of these objects is typically far smaller than the instrument psf and therefore all of them can be safely considered as \emph{point sources}, with observed profiles that are basically equal to the observing beam. In that case, we can effectively forget about the scale $R$ and assume that is something intrinsic to the profile $\tau_0$ that does not require to be made explicit.

But this cannot be the case for galaxy clusters: many of them are resolved objects even at the relatively low angular resolutions of \emph{Planck} \cite{ERCSC}. Detection algorithms must therefore provide not only a list of positions and intensities, but also of sizes of clusters. Moreover, they should give accurate photometry (values of $I_*$) for all the range of possible object sizes. MMF can be adapted to do precisely this in a very simple manner.

Let us assume that all the compact sources in a given multi-channel image have the same spatial profile except for a scale parameter $R$. For example we may consider the universal galaxy cluster pressure profile given by~\cite{arnaud10} with a different contrast radius $R_{500}$ for each cluster in our sample. If the $j^{\mathrm{th}}$ cluster has a true value of its scale parameter equal to $R_j$, then is it straightforward to prove that the filters (\ref{MMF}) satisfy the conditions of unbiasedness and maximum efficiency if and only if their scale is $S=R_j$. Moreover, as a part of the demonstration it can be shown that the signal-to-noise boosting given by the filter for that particular cluster is maximum also if and only if $S=R_j$. This suggests a very simple detection/estimation algorithm:
\begin{enumerate}
\item Filter the multi-channel image with a set of filters (\ref{MMF}) with different scales $S_i$.
\item Select as cluster candidates the positions of the maxima of the filtered images above a certain threshold.
\item For each cluster candidate $j$, obtain the curve showing the signal-to-noise boosting versus the scale $S_i$. Find the location $S_{max}(j)$ of that curve.
\item Make $R_j = S_{max}(j)$.
\end{enumerate}

\subsection{Unbiased matched multi-filters}

Galaxy clusters interact with CMB photons not only through the thermal SZ effect, but also through the so-called kinematic SZ effect due to the proper motion of the cluster. This is a perfect example of a case in which the same object produces two signals with identical spatial distribution but different spectral behaviours. The mixing of the two signals can affect negatively the determination of each of them. Normally the kinematic effect is at least one order of magnitude fainter than the tSZ and therefore the bias introduced by it in tSZ measurements can be  neglected. But the contrary is not true: the tSZ can contaminate significantly the estimation of the kinematic effect.  In order to avoid this~\cite{herr05} proposed a modification of MMF specifically tailored to cancel out this bias.

Let us consider a case in which we wish to observe the kinematic SZ effect without being disturbed by tSZ. In thermodynamic units the vector $\mathbf{f}=[f_1,\ldots,f_N]$ of the thermal effect is given by equation (\ref{eq:tSZ}). The spectral behaviour of the kinematic effect is, in the same units, flat: $[1,\ldots,1]$. We look for a set of $N$ filters $\Phi_i,\ldots,\Phi_N$ such that the estimation of the kinematic effect is not affected by the thermal. A sufficient condition for this is
\begin{eqnarray}
\int d\vec{k} \ \boldsymbol{\tau}^t \mathbf{\Phi} & = & 1 \nonumber \\
\int d\vec{k} \ \mathbf{F}^t \mathbf{\Phi} & = & 0, \\
\end{eqnarray}
\noindent
which can be interpreted as a kind of orthogonality with respect to the spectral behaviour laws of the components.
The solution, when we add the maximum efficiency constraint, looks similar to the MMF:
\begin{equation}
\mathbf{\Phi} = \frac{1}{\Delta} \mathbf{P}^{-1} \left( - \beta \mathbf{F} + \alpha \boldsymbol{\tau} \right).
\end{equation}
\noindent
In this equation the constants $\alpha$, $\beta$ and $\Delta$ are given by
\begin{eqnarray}
\alpha & = & \int d\vec{k} \  \mathbf{F}^t \mathbf{P}^{-1} \mathbf{F}, \nonumber \\
\beta  & = & \int d\vec{k} \  \boldsymbol{\tau}^t \mathbf{P}^{-1} \mathbf{F}, \nonumber \\
\gamma & = & \int d\vec{k} \  \boldsymbol{\tau}^t \mathbf{P}^{-1} \boldsymbol{\tau}, \nonumber \\
\Delta & = & \alpha \gamma - \beta^2.
\end{eqnarray}
\noindent
Similarly, it is possible --albeit less interesting-- to design filters that give the thermal SZ effect while cancelling the bias introduced by the kinematic effect. Both families of filters are called \emph{unbiased matched multi-filters} (UMMF).

\subsection{Unknown spectral behaviour}

The prior knowledge of $\mathbf{f}$ is the fundamental key of the success of MMF and the reason why they are able to get high signal-to-noise ratio boosts with respect to the individual channels.
However, as we mentioned before this prior knowledge is only certain when considering the SZ effect (and while ignoring relativistic effects). Extragalactic radio and infrared sources are not that simple to model. One possible solution is to group sources in broad families --radio flat, radio steep, dusty galaxies of a certain type, etc-- and to define average spectral laws, such as (\ref{eq:powlaw}), for each family. This approach will not be optimal for individual sources and will certainly introduce biases in the photometry, but at least is expected to improve the number of detections with respect to single-channel detection.

Fortunately MMF allow us to solve the conundrum in a very elegant way by means of an adaptive filtering scheme similar to the one described in section~\ref{sec:sizes}. 

Now imagine that $\mathbf{f}$ describes the true (unknown) spectral behaviour of a source and that $\mathbf{g}=[g_i]$, $i=1,\ldots,N$ is a new vector of equal size as $\mathbf{f}$, but whose elements can take any possible value. We can define the MMF for vector $\mathbf{g}$: 
\begin{eqnarray} \label{eq:MMFg}
\mathbf{\psi}_g  & = & \alpha_g \ \mathbf{P}^{-1} \mathbf{G} \nonumber \\ 	
\alpha_g^{-1} & = & \int d \vec{k} \ \mathbf{G}^t \mathbf{P}^{-1} \mathbf{G} \nonumber \\
G_i & = & g_i \tau_i,
\end{eqnarray} 
\noindent
where all the definitions are in Fourier space and we have not written down the explicit dependence on $\vec{k}$ for simplicity. If we apply the filter (\ref{eq:MMFg}) to a source with true spectral behaviour $\mathbf{f}$ and true intensity $I_*$ we will get
\begin{equation}
I_g = \hat{D}(\vec{0}) = I_* \alpha_g \int d\vec{k} \ \mathbf{G}^t \mathbf{P}^{-1} \mathbf{F},
\end{equation}
\noindent
so $I_g \neq I_*$ unless $\mathbf{g} = \mathbf{f}$. This is no help since we do not know the true value $I_*$. But it can be shown~\cite{lanz10} that the quantity
\begin{equation} \label{eq:snrlanz}
SNR_g = \frac{I_g}{\sigma_g},
\end{equation}
\noindent
where $\sigma_g^2$ is the variance of the image filtered with (\ref{eq:MMFg}), is maximum when $\mathbf{g} = \mathbf{f}$. $SNR_g$ is obviously the signal-to-noise ratio of the source after filtering. This allows us to reproduce the same kind of algorithm that was used in section~\ref{sec:sizes} to detect the sources and estimate simultaneously their position, intensity and spectral behaviour (instead of their scale). The only difference here is that while in section ~\ref{sec:sizes} there was only one parameter per cluster to be determined (the scale), here we have $N-1$ parameters per cluster (components of $\mathbf{f}$) to determine. The method has been recently applied to the two highest frequency channels of WMAP \cite{lanz11}, leading to a new catalogue of 157 objects detected at the $5\sigma$ level at 94 GHz, which is a substantial improvement over the WMAP Five-Band Point Source Catalogue.

\section{Bayesian multichannel detection and estimation} \label{sec:bayes}

All the multi-channel detection methods above have a common two-step methodological approach: in a first moment the data are somehow pre-processed and then objects are detected by means of some criterion --typically thresholding-- that hopefully separates them from the background noise. For the pre-processing step, one chooses beforehand a given tool (e.g. linear combinations of channels, or a given kind of filter/wavelet) and adjusts a small number of parameters (e.g. the weights of the linear combination,  the scale of the wavelet, etc) according to some optimality criterion (e.g. unbiasedness, maximum efficiency). For the detection step, an arbitrary threshold is chosen with the hope to reach a compromise between minimizing the number of spurious detections and maximizing the number of true detections. A third step for the estimation of a number of parameters of the sources (e.g. intensity, size...) can be attempted after detection or, in some cases, simultaneously to it\footnote{Even if \emph{detection} and \emph{estimation} can be done at the same time under certain circumstances, it must be noted that in statistics they are fundamentally different concepts.} (as it occurred with MMF). Errors to the estimated parameters are calculated by means of some prescription such as Fisher analysis. All these steps are somewhat arbitrary and are focused only in the statistical properties of the background (and the deterministic spectral behaviour of the sources, in the case of multi-channel detection). This approach leaves out all probabilistic (and potentially useful)  information about how many sources are expected to be found above a given flux limit, how are they distributed as a function of intensity, how many classes of sources are there and in which proportion are they present in the data, etc.

The Bayesian system of inference is the only one that provides a
consistent extension of deductive logic to a
broader class of ‘degrees-of-belief’ by mapping them into the real
interval $[0,1]$ \cite{jaynes2003}. Bayesian inference provides a logically consistent way of tackling the detection problem as a part of the \emph{decision theory}, while incorporating a probabilistic description of the sources as \emph{a priori} information in a natural way. The Bayesian framework also provides a sensible detection criterion through the Bayesian posterior odds ratio \cite{jaynes2003,carvalho12}. In estimation, Bayesian methods give a full description of the a posteriori distribution of the parameters given the current data, thus allowing us to obtain expectation values, confidence level contours and any other statistics of interest. 

\newcommand{\TT}{\boldsymbol{\Theta}}
\newcommand{\DD}{\mathbf{D}}
A detailed description of multi-channel Bayesian detection of compact sources is given in \cite{carvalho12}. Here we will just summarize the main theoretical aspects of the problem. Let us start with Bayes' theorem
\begin{equation}
Pr \left( \boldsymbol{\Theta} | \mathbf{D}, H \right) =
\frac{ Pr \left( \mathbf{D} | \boldsymbol{\Theta},H\right) Pr \left( \boldsymbol{\Theta}  | H \right)}
{Pr \left( \mathbf{D} | H \right)}, \label{eq:BI_Params}
\end{equation}
\noindent 
where $\DD$ is the vector of the observations as in eq. (\ref{eq:model1}) and  is the vector $\TT$ contains all the relevant parameters (positions, intensities, sizes, etc) to the detection problem and $H$ is the underlining hypothesis. In the usual Bayesian terminology, $Pr(\TT | \DD , H )$ is the \emph{posterior}
probability distribution of the parameters, $Pr(\mathbf{D} | \boldsymbol{\Theta},H) \equiv \mathcal{L}(\boldsymbol{\Theta})$
is the \emph{Likelihood}, $Pr \left( \boldsymbol{\Theta}  | H \right) \equiv Pr(\boldsymbol{\Theta})$ is the \emph{prior} and
$Pr \left( \DD  | H \right) \equiv \mathcal{Z}$ is the \emph{Bayesian evidence}. The most simple detection scenario can be described as a decision between two incompatible hypothesis $H_1$ (there is a source) and $H_0$ (there is not a source). In this case, it can be shown \cite{jaynes2003,carvalho12} that the decision criterion that minimizes the expected loss (that is, that optimizes the balance between the number of true and false detections) is given by the following condition on the Bayesian odds ratio
\begin{equation}
 \ln \left[ \frac{Pr(H_1 | \mathbf{D})}{Pr(H_0 | \mathbf{D})} \right] 
=  \ln \left[ \frac{\mathcal{Z}_1}{\mathcal{Z}_0} \, \frac{Pr(H_1)}{Pr(H_0)} \right] 
 \overset{H_1}{\underset{H_0}{\gtrless}} \xi,
\end{equation}
\noindent 
where $\xi$ is a threshold that is fixed for any given loss function. The choice of the loss function depends on the particular setting of the experiment; a common custom in microwave astronomy is to give identical weight to a spurious detection than to a missing true detection.

\subsection{Likelihood}

The form of the likelihood is determined by the statistical properties
of the generalised noise (background sky emission plus instrumental
noise) in each frequency channel. If the generalised noise is  statistically homogeneous
it is more convenient to work in Fourier space, since there are no
correlations between the Fourier modes of the generalised noise. Please note that although Galactic emission is not homogeneous, one can always operate locally in patches small enough to make the homogeneity assumption valid (as a first approximation).
It is also common to assume that both the background emission and
instrumental noise are Gaussian random fields. This is a very accurate
assumption for instrumental noise and the primordial CMB,
but more questionable for Galactic emission. Under the previous assumptions, the likelihood ratio between the hypotheses $H_1$ and $H_0$ can be written as
\begin{eqnarray}
\label{eq:LikeRatioFinal}
\ln\left[\frac{\mathcal{L}_{H_1}(\boldsymbol{\Theta})}{\mathcal{L}_{H_0}(\boldsymbol{\Theta})}\right]
& = & \sum_{\vec{k}}
{\mathbf{D}}^t(\vec{k})\boldsymbol{\mathbf{P}}^{-1}(\vec{k})
{\mathbf{s}}(\vec{k};\boldsymbol{\Theta}) \nonumber \\ && -
\frac{1}{2} \sum_{\vec{k}}
{\mathbf{s}}^t(\vec{k};\boldsymbol{\Theta})
\boldsymbol{\mathbf{P}}^{-1}(\vec{k}){\mathbf{s}}(\vec{k};\boldsymbol{\Theta}).
\end{eqnarray} 
\noindent
Note that the effect of the products $\mathbf{X}^t \mathbf{P}^{-1} \mathbf{Y}$, with $\mathbf{X}$ and $\mathbf{Y}$ generic vectors, is to project the multi-channel data into one single equivalent channel (or plane), as it happened in the MMF and UMMF. The similarities are more profound. It is shown in \cite{carvalho12} that, when the source term $\mathbf{s}$ is written as the sum of $N_s$ compact sources distributed across the image, the likelihood ratio (\ref{eq:LikeRatioFinal}) can be split into two parts:
\begin{itemize}
\item A sum of `auto-terms' each containing just the parameters corresponding to the $i^{\mathrm{th}}$ source ($i=1,\ldots,N_s$) 
\item A sum of `cross-terms' with mixed parameters corresponding to the $i^{\mathrm{th}}$, $j^{\mathrm{th}}$ sources ($i,j=1,\ldots,N_s$)
\end{itemize} 
The cross-term goes quickly to zero when the sources are well separated, but must be taken into account when source blending is frequent (crowded fields). It is worth noting that maximising the likelihood ratio (\ref{eq:LikeRatioFinal}),
in the absence of the cross-term (negligible source blending), with respect to the source
intensities $I_{*,j}$, leads precisely to the MMF described in section~\ref{sec:mmf}. This is the multi-channel generalization of the well-known result that matched filtering is the solution of the generalized maximum likelihood test (GLRT) for Gaussian noise and a single source \cite{kay,herr10}. 

\subsection{Priors}

If the data model provides a good description of the observed data
and the signal-to-noise ratio is high, then the likelihood will be very
strongly peaked around the true parameter values and the prior will
have little or no influence on the posterior distribution. At the faint
end of the source population, however, priors will inevitably play
an important role. Moreover, since for most cases in astronomy the
faint tail overwhelmingly dominates the population, the selection
of the priors becomes important and has to be addressed very carefully.
Physical (informative) priors are particularly useful when addressing the detection problem, whereas non-informative priors can be more adapted to the task of parameter estimation once the sources have been detected \cite{carvalho12}. For multi-channel detection, the list of priors to be considered must include

\subsubsection{Prior on the positions}
 For extragalactic sources, it is reasonable to assume a uniform prior. Clustering can be a problem, particularly for dusty galaxies observed at $\nu \geq 300$ GHz.
 \subsubsection{Prior on the number of sources} Following the same rationale of local uniformity, i.e no clustering,
the probability of finding $N_s$ objects (above a given flux limit) in a
sky patch follows a Poisson distribution 
\begin{equation}
\pi(N_s) = Pr(N_s | \lambda) = e^{-\lambda} ~ \frac{\lambda^{N_s}}{N_s!},
\end{equation}
\noindent
where $\lambda$ is the expected number of sources per patch. 
\subsubsection{Prior on intensity}
A usual informative prior on the distribution in intensity of extragalactic sources is a Generalized Cauchy Distribution such as in \cite{argueso11}. This provides a good model for the observed
distribution of fluxes, fitting the de Zotti or Tucci models almost perfectly \cite{zotti05,tucci11}. Moreover, the distribution can be properly normalised
as required for evidence evaluation. For the case of galaxy clusters, a power law distribution fits well the  cluster
populations assuming a standard WMAP best-fit $\Lambda$CDM cosmology
\cite{komatsu11} and the Jenkins mass function \cite{jenkins01}.
\subsubsection{Prior on the sizes}
Point sources are best modelled by imposing
the prior $\pi(R)=\delta(R)$ on the ‘radius’. For galaxy clusters, the fact that a significant number of them can be resolved must be taken into account. In \cite{carvalho12} a truncated exponential law is found to fit the simulated catalogues very well.
\subsubsection{Prior on the spectral parameters}
For galaxy clusters observed through the thermal Sunyaev-Zel'dovich effect, the spectral behaviour can be safely fixed to (\ref{eq:tSZ}), if we ignore relativistic effects. For point sources, however, the fact that each source has a spectral behaviour that is different from all the others must be accommodated. Either if we use power laws such as (\ref{eq:powlaw}) for radio sources or grey body laws for dusty galaxies, there is at least one spectral parameter that changes from one source to another.  Informative priors can be derived from the available source counts models \cite{zotti05,tucci11}, or uniform priors can be used when physical models are not available. 
\subsubsection{Priors on the models}
If there is more than a kind of source in the data (e.g. we have radio sources, dusty galaxies, Galactic compact sources and galaxy clusters in the same image) then appropriate priors in the probability of the different hypotheses $H_j$ should be included in the Bayesian framework. More importantly, the Bayesian framework allows us to explicitly consider the probability of background fluctuations above a certain level in order to optimize the decision between the hypotheses $H_j$ and the null hypothesis $H_0$ (i.e. no source).  

\subsection{Bayesian detection methods in the literature}

Solving the Bayesian equations for object detection and parameter estimation typically implies sampling from a very complex posterior distribution
with variable dimensionality (dependent on the number of
objects). Thus the main problem that Bayesian methods need to overcome is the computational burden of evaluating integrals over the posterior and its marginals. This can be avoided in some cases by semi-analytical Maximum A Posteriori evaluation \cite{argueso11}, but if a full analysis of the posterior is required then sampling is unavoidable. Typical implementations include Monte-Carlo Markov chain
(MCMC) algorithms \cite{hobson03} and posterior refinements such as nested sampling \cite{feroz08,feroz09}. \cite{carvalho09} implemented a simultaneous multiple minimization code based on Powell's direction set algorithm \cite{NRF}, that is generalized to the multi-channel case in \cite{carvalho12}. A full description of that algorithm and its catalogue making procedure it out of the scope of this review.

\section{Polarized sources} \label{sec:pol}

An interesting case of multi-channel setting is the detection of extragalactic sources in polarization data. Mathematically the problem is not different from the general model given in section~\ref{sec:model}, but the particularities of polarization observations have motivated the appearance of some specific techniques apart from the ones already discussed.

CMB polarization has been described as the next observational
frontier of cosmology. In the
last few years, some methods have been specifically developed to
address the important problem of detecting compact sources in polarization data. In this section, we will give a
brief review of these methods, which have been applied to CMB
simulations as well as
to real data \cite{argueso09,caniego09,argueso11p}.

Polarization of light is conveniently described in terms of the
Stokes parameters $I$, $Q$, $U$ and $V$, (see~\cite{kamion} for an
excellent review on CMB polarization). $Q$ and $U$ are the linear
polarization parameters and $V$ indicates the circular polarization.
Whereas $Q$, $U$ and $V$ depend on the orientation of the receivers, the
total polarization, defined as
\begin{equation} \label{eq:P}
P \equiv \sqrt{Q^2+U^2+V^2},
\end{equation}
\noindent is invariant with respect to the relative orientation of
the receivers and the direction of the incoming signal, and
therefore has a clear physical meaning. This quantity can be treated
as the modulus of a vector. Thus, the methods presented in~\cite{argueso09} 
have to do with the study of a set of images which contain
signals whose individual intensities can be considered as components
of a vector, but where the quantity of interest is the modulus.
First, we will consider the case of linear polarization, $V\equiv
0$, given that the CMB is not circularly polarized in the standard
cosmological models~\cite{kamion}. However, since some models
predict a possible circular polarization~\cite{cooray03}, we will
comment briefly on this case later.

If we have a compact source embedded in the data of $ Q$ and $U$,
these can be expressed in the following way
\begin{equation} \label{eq:datamodel}
  D_{Q,U}(\vec{x}) = A_{Q,U}\tau (\vec{x}) + n_{Q,U}(\vec{x}).
\end{equation}
\noindent with $A_{Q,U}$ the compact source intensity in $Q$ and
$U$, $\tau (\vec{x})$, the beam profile and $n_{Q,U}(\vec{x})$ the
corresponding noise in both components. The $P$-map,
$P(\vec{x})\equiv (D_{Q}^2(\vec{x})+D_{U}^2(\vec{x}))^{1/2}$, is
characterised by a source at the centre of the image with amplitude
$A\equiv (A_Q^2+A_U^2)^{1/2}$ immersed in non-additive noise which
is correlated with the signal.

 We assume the same beam profile for the images in $Q$ and $U$, as
well as a Gaussian independently distributed noise with zero mean
and the same variance for $Q$ and $U$. Given these typical
conditions, the distribution of $P$ if a source is present is the
Rice distribution
\begin{equation} \label{eq:rice-p}
  f(P|A) = \frac{P}{\sigma^2}e^{-(A^2 + P^2)/2\sigma^2}
  I_0\left(A\frac{P}{\sigma^2}\right),
\end{equation}
\noindent where $\sigma$ is the noise rms deviation and $I_0$ is the
modified Bessel function of zero order. By using the Neyman-Pearson
lemma~\cite{herr10}, a filter is obtained which produces the maximum likelihood
estimator of the amplitude, $\hat{A}$. In the case of a pixelized
image, we can write
\begin{equation} \label{eq:mle}
  \hat{A} \sum_i\frac{\tau_i^2}{\sigma_i^2} =
  \sum_iy_i\frac{I_1(\hat{A}y_i)}{I_0(\hat{A}y_i)},\ \ \ y_i\equiv
  \frac{P_i\tau_i}{\sigma_i^2}.
\end{equation}
\noindent where the indexes refer to the pixels and $I_1$ the
modified Bessel function of the first order. This filter is
called the Neyman-Pearson filter (NPF). Alternatively, a matched
filter can be applied to each image and then, with the two filtered
images $Q_{MF}, U_{MF}$ a non-linear fusion $P \equiv
(Q_{MF}^2+U_{MF}^2)^{1/2}$ can be made pixel by pixel. This filter
is called the filtered fusion (FF).

Simulations with the \emph{Planck} characteristics showed that the FF
performed better than the NPF especially for low polarization
fluxes. The FF outperformed the NPF when the power of the detections
with a given significance and the flux and position estimation were
compared.

If the circular polarization is taken into account, the NPF and the
FF can be also calculated. Indeed, the filters can be computed for
the modulus of a vector with any number of components~\cite{argueso11p}. 
Simulations showed that in the case of circular polarization,
the FF also produces better results than the NPF.

Taking into account the results with simulations, \cite{caniego09} 
applied the FF to the detection of polarized sources in the
WMAP 5-year data. They detected, with a significance level greater
than $ 99.99 \%$ in at least one WMAP channel, 22 objects, 5 of
which were doubtful, since they did not have a plausible low
frequency counterpart. These detections were a clear advance with
respect to the $5$ polarized sources listed by the WMAP team when
they analysed the same data. The application of a filter targeted
specifically for polarization detection had a clear influence on the
improvement.

The application of these methods to Planck data can be expected in the future. 
It would also be interesting to combine them
with a Bayesian approach, so that some previous information about
polarization properties of the sources could be taken into account.

\section{Final remarks}

CMB image processing is a relatively young area of research.
Much work remains to be done. There is an incipient but
resolved interest among CMB cosmologists to incorporate the
newest ideas to solving the problem of compact sources in
microwave images.

In this paper we have reviewed the status of the different algorithms and methods that attempt the detection of extragalactic compact sources (galaxies and clusters of galaxies) using multi-channel (that is, obtained by means of more than one single detector) data. This should not be confounded with the traditional band-merging approach to catalogue making. We included the case of detection in polarized data because it is formally equivalent to the detection in multi-frequency experiments.  The techniques we have reviewed include linear fusion of images, different multi-channel filtering methods and Bayesian object detection. Although the area of research is young, all of these methods have already been applied to astronomical data sets with great success. In next years, however, we expect to see new interesting developments in the field.

One of the most promising ideas for future research is related to the notions of
sparsity and $l_p$-approximations. For the particular case of point like
objects, the idea of sparse dictionaries comes naturally \cite{spars09}.
However, the full application to multi-channel CMB compact source detection
has not been addressed yet. Wavelet techniques, that are very popular in single-frequency source detection, open another interesting possibility that has not been explored yet. The same applies to other space-scale representations such as Gabor and Wigner-Ville transforms. Undoubtedly, both Bayesian and multi-filtering techniques will see substantial improvements in the next few years. All these new developments will be really come in handy for the new generation of many channel cosmological surveys such as the upcoming J-PAS\footnote{http://w3.iaa.es/~benitez/jpas/survey.html}.


%


\section*{Acknowledgement}

The authors would like to thank the editors of this Special Issue for their invitation to submit a contribution. We also thank Prof. Gianfranco De Zotti for his useful advice. DH and FA acknowledge partial financial support from the Spanish
Ministerio de Ciencia e Innovaci\'on project AYA2010-21766-C03-01. DH also acknowledges the Spanish Ministerio de Educaci\'on for a Jos\'e Castillejo' mobility grant with reference JC2010-0096 and the Astronomy Department at the Cavendish Laboratory for their hospitality during the elaboration of this paper. FA also wants to thank the Astronomy Department at the Cavendish Laboratory for their hospitality during two short stays in the summer of 2011. The authors would like to thank Mike Hobson for his very useful contributions and for the wonderful, although very sharp and clear, descriptions. P. Carvalho is supported by a Portuguese fellowship (ref: SFRH/BD/42366/2007) from the Funda\c{c}\~ao para a Ci\^encia e Tecnologia (FCT).

\appendix

\begin{table}[!h]
\centering
\begin{tabular}{ll}
   &  \\ 
 $\vec{x}$, $\vec{y}$ & position vector  \\ 
 $\vec{k}$ & wave vector (Fourier)  \\ 
  $N$ & number of channels  \\ 
 $D_j$ & observed data at the $j^{\mathrm{th}}$ channel \\ 
 $s_j$ & signal at the $j^{\mathrm{th}}$ channel  \\ 
 $n_j$ &  noise at the $j^{\mathrm{th}}$ channel \\ 
 $A_j$ &  source amplitude at the $j^{\mathrm{th}}$ channel \\ 
 $\tau_j$ & source profile at the $j^{\mathrm{th}}$ channel \\ 
 $\tau^0$ & intrinsic source profile (before smearing by the beam) \\  
 $b_j$ & beam point spread function at the $j^{\mathrm{th}}$ channel \\ 
 $R$ & scale parameter  \\ 
  $\mathbf{P}$ & cross-power spectrum matrix   \\ 
  $\mathbf{f}$ & spectral behaviour of the source  \\ 
  
 $\mathbf{\Psi}$,$\mathbf{\Phi}$ &  matrix (or vector) of filters \\ 
  $SNR$ & signal-to-noise ratio   \\ 
 $\boldsymbol{\Theta}$ &  vector of parameters (position, size, amplitude...) \\ 
  $I,Q,U,V$ & Stokes' polarization  parameters \\
  $P$ & total polarization \\
  $I_0$ & modified Bessel function of zero order \\
\end{tabular} 
\caption{List of symbols used in this review} \label{tb:list}
\end{table}

\bibliographystyle{IEEEtran}
\bibliography{HAC}

\end{document}